# The Response of an Elastic Splitter Plate Attached to a Cylinder to Laminar Pulsatile Flow

Anup Kundu[1], Atul K. Soti[1], Rajneesh Bhardwaj[2*] and Mark C. Thompson[3]

[1]IITB-Monash Research Academy, Indian Institute of Technology Bombay, Mumbai, 400076, India.

[2]Department of Mechanical Engineering, Indian Institute of Technology Bombay, Mumbai, 400076 India.

[3]Fluids Laboratory for Aeronautical and Industrial Research (FLAIR), Department of Mechanical and Aerospace Engineering, Monash University, Melbourne, VIC 3800, Australia.

*Corresponding author (Email: rajneesh.bhardwaj@iitb.ac.in,

Phone: +91 22 2576 7534, Fax: +91 22 2572 6875)

*Abstract*

The flow-induced deformation of a thin, elastic splitter plate attached to the rear of a circular cylinder and subjected to laminar pulsatile inflow is investigated. The cylinder and elastic splitter plate are contained within a narrow channel and the Reynolds number is mostly restricted to $Re = 100$, primarily covering the two-dimensional flow regime. An in-house fluid-structure interaction code is employed for simulations, which couples a sharp-interface immersed boundary method for the fluid dynamics with a finite-element method to treat the structural dynamics. The structural solver is implicitly (two-way) coupled with the flow solver using a partitioned approach. This implicit coupling ensures numerical stability at low structure-fluid density ratios. A power spectrum analysis of the time-varying plate displacement shows that the plate oscillates at more than a single frequency for pulsatile inflow, compared to a single frequency observed for steady inflow. The multiple frequencies obtained for the former case can be explained by *beating* between the applied and plate oscillatory signals. The plate attains a self-sustained time-periodic oscillation with a plateau amplitude in the case of steady flow, while the superimposition of pulsatile inflow with induced plate oscillation affects the plateau amplitude. Lock-in of the plate oscillation with the pulsatile inflow occurs at a forcing frequency that is twice of the plate natural frequency in a particular mode and this mode depends on the plate length. The plate displacement as well as pressure drag increases at the lock-in condition. The percentage change in the maximum plate displacement, and skin-friction and pressure drag coefficients on the plate, due to pulsatile inflow is quantified. The non-linear dynamics of the plate and its coupling with the pulsatile flow are briefly discussed.

# 1 Introduction

Fluid-Structure Interaction (FSI) with large-scale flow-induced structure deformation has potential applications in complex biomedical as well as engineering flows. For instance, the non-linear dynamic response of a soft structure subjected to pulsatile flow is useful for understanding cardiac hemodynamics [1-4]. In addition, deforming thin structures are potentially useful for energy-harvesting devices, and recent studies have demonstrated thermal augmentation via flow-induced deformation of thin elastic plates [5-6]. In the following sub-sections, we review previous studies on flow past rigid as well as flexible structures.

## 1.1 Studies on pulsatile inflow past a rigid cylinder

Through investigations of the effect of pulsatile inflow perturbations on flow past a stationary cylinder, previous numerical studies have shown *lock-in* behavior (also referred to as phase-locking or synchronization), in which the vortex shedding frequency shifts to be commensurate with the pulsatile forcing frequency at inflow (see review by Griffin and Hall [7]). Measurements of the flow past a circular cylinder [8, 9] showed that lock-in occurred for pulsatile frequencies at approximately twice the vortex-shedding frequency. Meneghini and Bearman [10] plotted the lock-in range for different pulsatile frequencies and amplitudes. Guilmineau and Queutey [11] numerically studied the flow over an in-line oscillating cylinder with 20% oscillation amplitude of the cylinder diameter and for Reynolds number, $Re = 185$. They showed that the shed vortices switch from one side of the cylinder to the other, as the pulsatile flow frequency increases to a limiting value. Konstantinidis et al. [12] confirmed the lock-in characteristics observed in previous measurements [8, 9], and showed that the wavelength of the vortex street varies with the pulsatile flow frequency but the flow amplitude does not alter vortex spacing. More recently Leontini et al. [13] further quantified the lock-in behavior of a circular cylinder undergoing forced streamwise oscillations as a function of forcing frequency and amplitude, characterizing the wake response over a wide range of control parameters in the two-dimensional laminar regime.

## 1.2 Studies on steady/pulsatile inflow past flexible thin structures

In the context of steady inflow past flexible/elastic thin structures, previous studies documented the effects of the material properties of the structure and flow conditions on the response of the structure. While proposing a FSI benchmark for flow-induced deformation of elastic thin structures, Turek and Hron [14] showed that the flow past an elastic splitter plate attached to lee side of a rigid cylinder in two-dimensional laminar channel flow results in a self-sustained oscillation of the plate. Bhardwaj and Mittal [15] quantified the effect of Reynolds number, material properties and geometric non-linearity on the plate displacement as well as its frequency in the FSI benchmark proposed by Turek and Hron [14]. They showed that the oscillation frequency of the plate varies linearly with dilatational wave speed inside the plate (or its natural oscillation frequencies). Lee and You [16] showed that the plate length influences vibration modes of the splitter plate, and the plate displacement is a function of the Young's Modulus and its natural frequencies. Furquan and Mittal [18] investigated flow past two side-by-side square cylinders with flexible splitter plates and observed lock-in for the plate frequency closer to its first natural frequency. Shoele and Mittal [19] proposed stability curves for a flexible plate in an inviscid channel flow and showed that the plate oscillation frequency as well as its stability depends on the channel height. Shukla et al. [17] experimentally studied the effects of flexural rigidity as well as plate length on the response of the plate in the wake of a circular cylinder and found that the plate displacement collapses on a single curve for different cases of dimensionless bending stiffness.

The FSI of elastic, inextensible filaments attached to a cylinder was also reported in previous studies. Bagheri et al. [20] showed that a hinged flexible filament attached on a cylinder generates a net lift force without increasing drag on it, due to symmetry-breaking instability of the filament which oscillates in upper or lower part of the cylinder wake. Extending work of Bagheri et al. [20], Lacis et al. [21] showed the symmetry-breaking instability is similar to the instability of an inverted pendulum. Note that these studies considered very low values of structure-fluid density ratio (O(0.1)) as well as flexural rigidity (O(0.001)- O(0.1)), which is two-three orders of magnitude lesser than the values used in the present study. An attached filament on a cylinder also helps in reducing mean drag as well as fluctuations of lift on the cylinder, as reported by Wu et al. [22].

Very few studies are reported in the context of pulsatile inflow past flexible/elastic thin structures in the literature. For instance, Razzaq et al. [23] studied the FSI interaction of the elastic walls of an aneurysm with an implanted stent structure subjected to pulsatile flow. Habchi et al. [24] studied twin elastic thin plates mounted in cross-flow configuration at a distance and subjected to the pulsatile flow. They reported that the plates oscillate in opposite-phase and in-phase for larger and smaller value of Young's modulus of the plate, respectively.

### 1.3 Objectives of the present study

While the effect of the pulsatile flow on oscillating rigid structures is well-documented and understood, previous reports [23, 24] available for the pulsatile inflow past flexible, deformable structures did not investigate the effects of forcing frequency and flow amplitude. In addition, the response of the structure for different material properties is poorly reported. The non-linear interaction of the pulsatile flow with the moving structure leads to complex system behavior, such as lock-in and beating. Such effects have not been investigated to the best of our knowledge. The objective of the present study is to investigate the effect of the pulsatile flow on the flow-induced deformation of a thin, elastic structure for an inline flow configuration. To achieve this, the FSI benchmark case proposed by Turek and Hron [14] is extended to account for the pulsatile inlet flow, and is used to investigate the coupling of the forcing flow frequency as well as amplitude with the frequency of the oscillating plate. The FSI model employed to tackle this problem is discussed in section 2 and results are presented in section 3.

## 2 Computational Model

The FSI modeling with large-scale flow-induced structural deformation involves complex 3D geometries, moving structural boundaries in the fluid domain, and geometric and/or material nonlinearity of the structure. The coupling of the governing equations of the fluid with those of the structure brings additional non-linearity to the governing equations. In order to address these complexities, an in-house FSI solver is employed. The solver is based on a sharp-interface immersed boundary method, in which the governing equations of the flow domain are solved on a fixed Cartesian (Eulerian) grid while the movement of the immersed structure surfaces is tracked within a Lagrangian framework. As reviewed by Mittal and Iaccarino [25], the Immersed Boundary Method (IBM) is designed to treat 3D moving boundaries but using a simple and efficient Cartesian grid. Since the governing equations are solved on a body non-conformal Cartesian grid, there is no need for remeshing to treat deforming or moving structure boundaries in the fluid domain, provided spatial resolution is adequate. The FSI solver employed in the present study was developed by Mittal and co-workers [26 - 29], and later further developed for large-scale flow-induced deformation by Bhardwaj and Mittal [15]. The flow is governed by the unsteady, viscous, incompressible Navier-Stokes equations written in dimensionless form as follows,

$$\frac{\partial v_i}{\partial x_i} = 0, \tag{1}$$

$$\frac{\partial v_i}{\partial t} + \frac{\partial v_i v_j}{\partial x_j} = -\frac{\partial p}{\partial x_i} + \frac{1}{Re}\frac{\partial^2 v_i}{\partial x_j^2}, \tag{2}$$

where $i, j = 1, 2, 3$, and $v_i$, $t$, $p$ and $Re$ are velocity components, time, pressure, and Reynolds number based on mean flow velocity and cylinder diameter, respectively. These equations are discretized in space using a cell-centered, collocated (non-staggered) arrangement of primitive variables ($v_i$, $p$) using a second-order, central-difference scheme for all spatial derivatives. Because the computational methodology has been previously well-documented [15, 26 - 29], only a brief overview of the method is provided. The unsteady Navier-Stokes equations are marched in time using a fractional-step scheme [26, 30] that involves two sub-steps: solving an advection-diffusion equation, followed by a substep to calculate the pressure by solving a Poisson equation. During the first step, both the convective and viscous terms are treated (semi-) implicitly are using Crank-Nicolson scheme to improve the numerical stability and maintain second-order accuracy. In the second substep, the pressure Poisson equation is solved with the constraint that the final velocity be divergence-free. Once the pressure is obtained, the velocity field is updated to its final divergence-free value. A fast geometric multi-grid solver [31] is used to solve the pressure Poisson equation. A sharp-interface immersed boundary method based on a multi-dimensional ghost-cell methodology is utilized to apply flow boundary conditions on the boundaries of the structure. Furthermore, the immersed structure boundary is represented using an unstructured grid of triangular elements within the flow domain.

An open-source finite-element-based structural dynamics solver – Tahoe[©] [32] – is employed. In the present study, the structure is modeled as Saint Venant-Kirchhoff material, suitable for large deformations. The constitutive relation between the stress and the strain is based on the Green-Lagrangian strain tensor $E$ and the second Piola-Kirchhoff stress tensor $S(E)$, which is a function of $E$. The second Piola-Kirchhoff stress tensor can be expressed in terms of the Cauchy stress tensor $\sigma$ as follows [33],

$$\mathbf{S} = J\mathbf{F}^{-1}\sigma\mathbf{F}^{-T} \qquad (3)$$

where $J$ is the determinant of the deformation gradient tensor $\mathbf{F}$. The Green-Lagrangian strain tensor $\mathbf{E}$ is defined as,

$$\mathbf{E} = \frac{1}{2}(\mathbf{F}^T\mathbf{F} - \mathbf{I}) \qquad (4)$$

The structural solver Tahoe© [32] is strongly coupled with the in-house flow solver using an implicit partitioned approach, described by Bhardwaj and Mittal [15]. The implicit coupling ensures numerical stability at low structure-fluid density ratio [15]. The boundary conditions at the fluid-structure interface are described as follows. No slip boundary conditions are applied for the velocity at the fluid-structure interface which represents continuity of the velocity at the interface,

$$v_{i,f} = \dot{d}_{i,s}, \qquad (5)$$

where subscripts $f$ and $s$ denote fluid and structure, respectively. In addition, the balance of forces is applied at the interface,

$$\sigma_{ij,f} n_j = \sigma_{ij,s} n_j, \qquad (6)$$

where $n_j$ is the local surface normal pointing outward from the surface. The pressure loading on the structure surface exposed to the fluid domain is calculated at the current location of the structure using interpolated fluid pressure via bilinear interpolation, as described by Mittal et al. [26].

Briefly, for the coupling, an outer iteration is performed at each time step until convergence is achieved between the flow and structural solver [15]. At each outer iteration, both the flow and structural dynamics are solved while updating the boundary conditions at the fluid-structure interface. Convergence of the coupled system is assumed when the calculated residual (measured by the $L_2$ norm of the displacement of the fluid-structure interface) drops below a user-defined value. Typically, 5-10 outer iterations are required to achieve convergence in the simulations reported in the present study.

## 2.1 Code validation

The flow solver has been extensively validated by Mittal et al. [26] against several benchmark problems, such as flow past a circular cylinder, sphere, airfoil, and a suddenly accelerated circular cylinder and a normal plate. In this section, additional qualitative and quantitative validations are performed for the flow past a circular cylinder with pulsatile inflow boundary condition applied at the channel inlet. The large-scale flow-induced deformation module in the in-house FSI solver was previously validated by Bhardwaj and Mittal [15], and this is briefly described here for completeness.

### 2.1.1 Pulsatile inflow past a stationary cylinder

Vortical wake structures for pulsatile flow past a circular cylinder placed in a channel are qualitatively compared. This flow problem was considered previously by Al-Sumaily and Thompson [34]. The computational domain is $23D \times 4D$, where $D$ is the cylinder diameter, and the center of the cylinder is positioned at ($8D$, $2D$). A fully-developed dimensionless pulsating flow velocity $u(t)$ is applied at the inlet of the channel and is expressed as follows [34],

$$u = 1 + A_\text{f} \sin(2\pi \text{St}\, t), \tag{7}$$

where $A_\text{f}$, $St$ and $t$ are amplitude of the pulsatile inflow, Strouhal number and time respectively. No-slip boundary conditions at the top and bottom walls, and Neumann outflow boundary condition at the right boundary are enforced for the simulation. The computed vorticity contours obtained by using the same parameters in our FSI model ($St_\text{f} = 0.8$) are plotted in the left column of Fig. 1. These contours are compared with those of Al-Sumaily and Thompson [34] plotted in the right column of this figure, at a series of different time instances. A good agreement in the vorticity field predictions is observed, notably, in terms of both vortex strength and shape, thereby providing confidence in the implementation of oscillating flow boundary conditions.

Quantitatively the flow model is validated for predicting the flow past a circular cylinder under steady flow and pulsatile flow with non-zero mean velocity [35, 36]. The computational domain is $16D \times 30.5D$ where $D$ is cylinder diameter, and the center of the cylinder is at ($8D$, $8D$). The dimensionless inflow velocity in the channel is expressed as,

$$u = 1 + A_\text{f} \cos(2\pi f_\text{f} t), \tag{8}$$

where $A_\text{f}$, $f_\text{f}$ and $t$ are flow amplitude, frequency of the pulsatile inflow and time, respectively. A good agreement is obtained for predictions of maximum lift coefficient, average drag coefficient and Strouhal number for the steady flow ($A_\text{f} = 0$), as given in the Table 1. Results for the pulsating flow past a stationary cylinder were compared with those of Li et al. [35]. The forcing frequency was taken as natural vortex shedding frequency ($f_\text{f} = 0.1639$). The first and second frequency of the lift coefficient signals, obtained by FFT analysis, at different forcing amplitudes are given in Table 2. The comparisons between the frequencies are good and validates our flow model.

## 2.1.2 Large-scale flow-induced deformation

The large-scale flow-induced deformation module as part of the in-house FSI code was validated by Bhardwaj and Mittal [15] against the FSI benchmark problem proposed by Turek and Hron [14]. In this benchmark problem, a cylinder with a $3.5D \times 0.2D$ thin elastic plate with specified material properties attached at its rear is placed inside a channel of width $4.1D$, where $D$ is the cylinder diameter (Fig. 2A). The fluid is taken to be Newtonian and incompressible. The plate is considered to consist of Saint Venant-Kirchhoff material, which accounts for geometric nonlinearity for a linear elastic material [33]. The boundary conditions for the benchmark problem are illustrated in Fig. 2A. No-slip boundary conditions are applied at the channel walls and immersed structure boundary. Zero Neumann boundary condition is applied for the velocity at the channel outlet. At the inlet, a fully developed, parabolic, steady velocity profile is applied, expressed in dimensionless form as follows [14],

$$u_{\text{steady}} = 6u_m \left(\frac{y}{H}\right)\left(1-\frac{y}{H}\right), \tag{9}$$

such that the dimensionless mean velocity in the channel of dimensionless height $H$ is $u_m$. The length and velocity scale used for non-dimensionalization are cylinder diameter ($D^*$) and mean velocity ($u_m^*$), respectively. Note that the superscript $^*$ and subscript m denote dimensional variable and mean value, respectively. The Reynolds number ($Re$) and dimensionless Young's modulus ($E$) are defined as follows,

$$Re = \frac{\rho_f^* u_m^* D^*}{\mu^*}, \tag{10}$$

$$E = \frac{E^*}{\rho_f^* u_m^{*2}}, \tag{11}$$

where $\rho_f^*, \mu^*, E^*$ are dimensional fluid density, dynamic viscosity and Young's modulus, respectively, and subscript f denotes fluid. The dimensionless structure density is expressed as,

$$\rho_s = \frac{\rho_s^*}{\rho_f^*} \tag{12}$$

where subscript s denotes structure. In the FSI benchmark, the following values are considered for the simulation setup [14]: $u_m = 1$, $D = 1$, $Re = 100$, $E = 1.4 \times 10^3$, $\rho_s = 10$. The Poisson's ratio is taken as 0.4 in the structural solver. The validation by Bhardwaj and Mittal [15] was conducted for the time-varying cross-stream position of the plate tip ($Y_{\text{tip}}$) and its oscillation frequency ($St_p$), after the plate has reached a self-sustained periodic oscillatory state. In Ref.

[15], $St_p$ was in excellent agreement while the difference in $Y_{tip}$ was around 11%, as compared to the benchmark data of Turek and Hron [14]. In the present work, we investigated the source of this difference and it is attributed to the following factors. In Ref. [15], the plate was considered with rounded corners (shown in the inset of Fig. 2A) and the simulation did not include contribution of shear force in traction boundary condition at the fluid-structure interface. These factors contributed around 8% and 3% difference with respect to the values of Turek and Hron [14], respectively. The time-varying $Y_{tip}$ and $X_{tip}$ obtained in the revised simulation performed in the present work are compared with the respective results in Refs. [14, 15] and are Fig 2B and Fig. 2C, respectively. The plate displacement as well as frequency is in excellent agreement with the published results of Turek and Hron [14] in the present work. The computed values of $St_p$ and $Y_{tip}$ are listed in Table 3 along with the values reported in the published studies [14-16, 18, 37].

## 3  Results and Discussion

In this section, we investigate the effect of pulsatile inflow on the flow-induced deformation and associated wake flow for the flow past a cylinder with a thin trailing elastic plate attached (Fig. 2A). The boundary conditions are same as those described in section 2.1.2 and are shown in Fig. 2A, except at the left channel boundary. A fully-developed pulsatile inflow velocity at the left boundary is prescribed and is expressed in non-dimensional form as follows,

$$u_{pulsatile} = u_{steady}\left(1 + K\sin(2\pi St_f t)\right), \tag{13}$$

where $u_{steady}$, $St_f$ and $t$ are dimensionless steady component of the velocity given by eq. 9, Strouhal number and time, respectively. The parameters used for the non-dimensionalization for eq. 13 are same as the ones used in section 2.1.2. The inflow velocity in eq. 13 is superimposition of the steady inflow ($u_{steady}$) and pulsatile inflow ($u_{steady} K\sin(2\pi St_f t)$), where $K$ is a constant in range of [0, 1] and controls the fraction of the unsteady component in the total inflow. A typical mesh used in the simulation in the present study is shown in Fig. 4. A non-uniform Cartesian grid with stretching is employed in the computational domain as shown in Fig 4A. Zoomed-in views of the grid in the vicinity of the immersed boundary and downstream are shown in Fig 4B and 4C, respectively. A uniform grid is used in the region in which the plate is expected to move and non-uniform grid stretching is used from this region to the wall (Fig. 4B). In the present section, we use the plate with rounded corners as shown in Fig. 4B. This remainder of this section is organized as follows:

- First, grid convergence and domain independence studies are presented in section 3.1.

- Second, the effect of plate length and structure-fluid density ratio is discussed on the plate oscillation in case of steady state inflow in section 3.2.
- Third, the effect of the pulsatile flow frequency ($St_f$) on the flow-induced deformation and associated flow fields are compared in section 3.3. The flow frequency is varied between 0.1 and 1.0, while keeping the flow amplitude constant ($K = 0.4$).
- Fourth, the effect of the flow amplitude ($K$) on the plate dynamics and flow fields are investigated in section 3.4. The flow amplitude ($K$) is varied between 0.0 and 1.0, while keeping the flow frequency constant ($St_f = 0.4$).
- Finally, a parameter map is presented in section 3.5 to specify the lock-in region and plate response based on the data obtained from several simulation sequences.

## 3.1 Grid and domain size convergence study

We performed grid convergence study with three different non-uniform Cartesian grids, $256 \times 160$, $384 \times 160$ and $480 \times 224$, under pulsatile inflow ($K= 0.4$, $St_f = 0.2$) for flow past an elastic splitter plate behind a cylinder. The time-step was set to $\Delta t = 0.01$. All other simulation setup parameters are given in section 2.1.2. The minimum grid sizes in $x$ and $y$ direction, $\Delta x_{min}$ and $\Delta y_{min}$ respectively, are listed in Table 4. The tip deflection ($Y_{tip}$) signals obtained for the different grids are compared in Fig. 5A, with the inset showing the minor differences observed in peak amplitude for the different grids. The errors with respect to the finest grid are listed in Table 4. Since the relative error for $384 \times 160$ grid, as shown in Table 4, is one order of magnitude smaller than that for $256 \times 160$ grid, the $384 \times 160$ grid ($\Delta x_{min} = \Delta y_{min} = 0.02$) was selected for all simulations presented in present work.

The domain independence study was conducted with four domains of sizes $20D \times 4.1D$, $25D \times 4.1D$, $30D \times 4.1D$, $40D \times 4.1D$. The steady inflow was considered and simulation parameters are given in section 2.1.2. The tip displacement ($Y_{tip}$) signals obtained from four different domains considered are compared in Fig. 5B. The inset of Fig. 5B shows the minor differences observed in peak amplitude for the different domains considered. The error with respect to the longest domain considered are listed in Table 5. Since the error for $20D \times 4.1D$ domain is lesser than 1%, this domain size is considered for all simulations presented in this paper.

## 3.2 Steady inflow

In this section, we investigate the effect of problem parameters on the plate oscillation frequency in FSI benchmark described in section 2.1.2. We vary Reynolds number (*Re*), structure-fluid density ratio ($\rho_s$) and plate length ($L$), keeping all other parameters same in cases 2, 3 and 4 listed in Table 6, respectively. The maximum computed displacement ($Y_{tip}$) and oscillation frequency ($St_p$) in all cases are summarized in Table 6. The natural frequencies ($St_{ni}$) of the plate in first three modes calculated using the modal analysis are also listed in Table 6. Using Euler–Bernoulli beam model, the natural frequency ($St_{ni}$) of the vibration of a cantilever beam in dimensionless form is given by [18, 38],

$$St_{ni} = \frac{k_i^2}{2\pi}\sqrt{\frac{EI}{\rho_s A L^4}} \qquad (14)$$

where $EI$ is the dimensionless flexural rigidity of the beam, $i = 1, 2, 3$ represents frequency modes of the plate and $k$ is the respective constant for the modes. $\rho_s$, $A$ and $L$ are dimensionless density, cross-sectional area and length of the plate respectively. The values of $k$ are 1.875, 4.694 and 7.855 for first, second and third mode of the natural frequency, respectively.

In the FSI benchmark (case 1 in Table 6, $L = 3.5D$), the plate initially exhibits small deformation and reaches a periodic self-sustained oscillation with a constant amplitude, displaying a sinusoidal-like wave pattern[1] of the time-varying displacement of the tip of the plate [15] (Fig. 2B). The superimposed shapes of the deformed plate at different time instances are shown in Fig. 3A(left). In this case, the plate oscillation frequency is close to the natural frequency in the second mode ($St_p \cong St_{n2}$, case 1 in Table 6). In cases 2 and 3 in Table 6, we vary $Re = 200$ and $\rho_s = 5$, respectively, keeping all parameters same and the simulated plate frequency $St_p$ is again close to the second mode of the natural frequency ($St_p \cong St_{n2}$). In case 4, we vary plate length to $L = D$ and the plate oscillates with the frequency close to the first mode of natural frequency ($St_p \cong St_{n1}$). The superimposed shapes of the deformed plate at different time instances for this case are shown in Fig. 3B.

Thus, the plates with length $L = 3.5D$ and $L = D$ oscillate with second and first mode of the natural frequency, respectively. This can be explained by the fact that the pressure loading on the longer plate is non-uniformly distributed as compared to that on the shorter plate. The non-uniform loading results in larger bending and consequently the plate vibration is closer to that in the second mode. A qualitative comparison of the pressure distribution at the instance of maximum plate deformation is shown in Fig. 3 (right column) and confirms this hypothesis. Note that these observations are consistent with those given by Lee and You [16].

---

[1] See also supplementary computer animation, S1.avi

### 3.3 Effect of pulsatile inflow frequency

In this section, we discuss the effect of pulsatile inflow on the plate oscillation. In pulsatile inflow case, the forcing frequency interacts with the natural oscillation frequency of the plate, which results in beating and lock-in signals. The effect of the pulsatile inflow frequency ($St_f$) is studied by varying it within the range [0.0, 1.0], while keeping the flow amplitude constant at $K = 0.4$. The time-varying tip displacement of the plate ($Y_{tip}$) is shown in the left column of Fig. 6. The power spectra of these signals are shown in the middle column, indicating the dominant frequencies in the signals. Note that the difference in the applied flow frequency ($St_f$) and the oscillating plate frequency ($St_{n2}$) generates *beating* between these frequencies, clearly evident in the tip displacement evolutions. For instance, at $St_f = 0.1, 0.2$ and $0.4$, the plate oscillates with $|St_f \pm kSt_{n2}|$ and $kSt_{n2}$ ($k$ = integer), as shown in the power spectra, plotted in middle column of Fig. 6. Lock-in condition occurs for $St_f = 0.4$, for which $St_f \sim 2St_{n2}$, which results in the largest plate displacement[2]. For the lower frequencies $St_f = 0.1 - 0.5$, the vorticity contours at the instance of maximum plate deformation, plotted in the right column of Fig. 6 show that the shear layers at the top and bottom of the plate roll up to form strong positive and negative vortices, resepctively. At higher frequencies ($St_f > 0.5$), the shear layer shows the formation of the two or more smaller vortices of the same sign on each side of the plate, showing the strong effect of pulsatile flow on the wake.

We quantify the effect of pulsatile flow in terms of the drag coefficients for the *plate*, defined as,

$$C_{DP} = \frac{2F_{DP}^*}{\rho_f^* u_m^{2*} D^*} \quad (15)$$

$$C_{DS} = \frac{2F_{DS}^*}{\rho_f^* U_m^{2*} D^*} \quad (16)$$

$$C_D = C_{DP} + C_{DS} \quad (17)$$

where $C_{DP}$, $C_{DS}$ and $C_D$ are pressure, skin-friction and total drag coefficient of the plate, respectively. $F_{DP}^*$ and $F_{DS}^*$ are the pressure and shear force per unit span-wise length on the plate, respectively.

In order to quantify the influence of pulsatile inflow on the drag on the plate and its components, we plot the time-variation of $Y_{tip}$, $C_{DP}$, $C_{DS}$, and $C_D$ for steady inflow and two cases of pulsatile inflow in Fig 7A and 7B, C, respectively. The vorticity distribution at different time instances is shown in the insets. The plots of the pulsatile inflow are presented for lock-in condition ($K = 0.4$, $St_f = 0.4$) in Fig. 7B. In Fig. 7A for the steady inflow, the maximum pressure drag as well as skin friction drag occurs at the maximum plate displacement (at $t \sim 186$ and $189$). The contribution of the skin-friction drag in the total drag is around 13% at these instances. The maximum pressure drag is attributed to blockage of the flow created by the deformed plate in the channel at the instance of the maximum deformation. As expected, the total drag and its components are the lowest at the instance of the mean position of the tip ($t \sim 187.5$).

---

[2] See also supplementary computer animation, S2.avi

In case of the pulsatile inflow (Fig 7B), the maximum pressure drag as well as skin friction drag also occurs at the maximum plate displacement (at $t \sim 187.5$ and $190$). However, the contribution of the skin friction drag in the total drag is 33%, around three times larger than that in the case of the steady inflow in Fig 7A. Interestingly, the skin friction drag is negative at the instance of the mean position of the tip ($t \sim 189$). This observation may be attributed to the formation of shear layers along the plate length due to the pulsatile inflow and is described as follows. As shown in the inset of Fig 7B, at the instance of the maximum tip displacement at $t \sim 187.5$ ($t \sim 190$), a shear layer of negative (positive) vorticity at the top (bottom) of the plate roll up to form strong negative (positive) vortex near the cylinder and another negative (positive) vortex which is about to shed in the downstream. On the other hand, at the mean position of the tip ($t \sim 189$), a shear layer of positive vorticity dominates along the plate length, which corresponds to negative skin friction drag at this instance.

Similarly, the time-variations of $Y_{tip}$, $C_{DP}$, $C_{DS}$, and $C_D$ at larger forcing frequency ($K = 0.4$, $St_f = 0.8$) are plotted in Fig 7C. Due to increased forcing frequency, the tendency to roll the vortices over the plate as well as their strength decreases (as seen in the insets in Fig 7C) and it results in larger pressure perturbations near the structure. Therefore, the maximum pressure drag at the instance of the maximum displacement at $St_f = 0.8$ is around 50% larger than that in $St_f = 0.4$.

To further quantify the effect of the pulsatile flow frequency, it is useful to define the percentage change in a flow quantity with respect to the steady inflow,

$$\Delta \eta = \frac{\eta_{pulsatile} - \eta_{steady}}{\eta_{steady}} \times 100\% \tag{18}$$

where $\eta_{pulsatile}$ and $\eta_{steady}$ are the flow quantities for the pulsatile and steady inflow, respectively. Fig. 8A plots $\Delta\eta$ for RMS values of $Y_{tip}$, $C_{DP}$, and $C_D$ as a function of pulsatile flow frequency, $St_f$. $\Delta\eta$ for $C_{DP,\,RMS}$ and $C_{D,\,RMS}$ scales non-monotonically with $Y_{tip,\,RMS}$ for $St_f \in [0.1, 0.5]$. The total drag and pressure drag contribution show a significant increase at the lock-in condition ($St_f \sim 2St_{n2}$), due to a 16 % larger plate displacement. Similarly, $C_{DP}$, and $C_D$ show decrease of around 15-25% in Fig. 8A due to decrease in RMS value of $Y_{tip}$ at $St_f = 0.2$ since $Y_{tip}$ shows a strong variation with time due to the beating in the Figure 6A (second row). At higher frequencies, $St_f \in [0.6, 1.0]$, $\Delta\eta$ for $C_{DP,\,RMS}$ is significantly larger (~25-40%) in Fig. 8A. As explained earlier, this is due to the decrease in strength of the rolling vortices over the plate (see insets of Fig 7C) which results in larger pressure perturbations near the structure and increases pressure drag. Thus, the pressure drag is larger for $St_f \in [0.6, 1.0]$ and the total drag also shows similar characteristics except at $St_f = 1.0$. The total RMS drag at $St_f = 1.0$ reaches to a value, comparable to that computed in the steady inflow case and is around 10% larger due to increase in the shear drag on the plate, explained in the following paragraph. The flow field at $St_f = 1.0$ becomes qualitatively similar to that in the steady inflow case, due to the decaying strength of the rolling vortices at the top and bottom of the plate. The vorticity field at the instance of the maximum displacement shown in the inset of Fig 6B (last row) is qualitatively similar to that for the steady inflow, in the inset of Fig. 7A, except in close proximity to the surface of the plate.

In order to quantify the contribution of the skin-friction drag, we define percentage of the skin-friction drag coefficient with respect to total drag coefficient $\Delta\psi$, as follows,

$$\Delta\psi = \frac{C_{DS,RMS}}{C_{D,RMS}} \times 100\% \tag{19}$$

As plotted in Fig 8B, $\Delta\psi$ is more than 30% for $St_f \in [0.2, 1.0]$. The increase in the skin-friction drag is attributed the formation of shear layers along the plate length due to pulsatile inflow, as explained earlier (see insets of Figs. 7B and 7C). A slight dip in $\Delta\psi$ at lock-in is explained by shedding of rolled vortices over the plate, which pushes the shear layer along the plate length[3]. Overall, the total drag is significantly larger (20%-50%) for the pulsatile flow cases as compared to the steady inflow case for $St_f \in [0.4, 1.0]$ and $K = 0.4$.

### 3.4 Effect of pulsatile inflow amplitude

In this section, the effect of the forcing flow amplitude on the flow-induced deformation of the splitter plate at constant pulsatile inflow frequency, $St_f = 0.4$, is examined. The time-varying plate displacement ($Y_{tip}$), power spectra of $Y_{tip}$ and vorticity contours at instance of maximum $Y_{tip}$ are plotted in the left, middle and right columns of Fig. 9, respectively. The flow amplitudes investigated are $K \in [0.0, 1.0]$ and $K = 0.0$, $St_f = 0.0$ corresponds to the steady inflow at the inlet. Since $St_f = 0.4$ corresponds to the lock-in frequency, as simulated in section 3.3, the beating is not observed in the simulated cases shown in Fig. 9, in contrast to many of the cases examined in section 3.3. Indeed, the power spectra of $Y_{tip}$, plotted in Fig. 9, show lock-in at all flow amplitudes. The vorticity contours are plotted in the right column of Fig. 9, showing that the vortices on the top and bottom sides of cylinder surface roll up increasingly tightly and are clearly identifiable as discrete entities as they move along the plate, as the forcing amplitude increases.

---

[3] See also supplementary computer animation, S2.avi

As in section 3.3, Fig. 10A shows the percentage change in various system characteristics with respect to the steady inflow: the maximum plate deformation ($Y_{tip, RMS}$), total drag ($C_{D, RMS}$) and pressure drag ($C_{DP, RMS}$) for the flow amplitudes tested for the lock-in condition. $C_{D, RMS}$ and $C_{DP, RMS}$ show almost a linear increase due to increase in $Y_{tip, RMS}$, implying increased blockage of the flow by the deformed plate in the channel at the instance of the maximum deformation. The maximum percentage increase in the total pressure drag ($C_{D, RMS}$) is approximately 75% and the plate deformation increases by 31%, both at $K = 1$.

The contribution of the skin friction drag with respect to total drag ($\Delta\psi$, eq. 19) with respect to flow amplitude $K$ is plotted in Fig. 10B. In general, we note a linear increase in $\Delta\psi$ with $K$ and the largest value 50% occurs at $K = 1$. The increase in shear drag with flow amplitude is due to the increasing strength of the rolling vortices and shear layers along the plate length. The vorticity signatures at the instance of the maximum plate deformation, plotted in the insets of Fig 9, verify this hypothesis. Overall, the total drag is significantly larger (20% to 75%) for the pulsatile flow cases as compared to the steady inflow case for $K \in [0.4, 1.0]$ and $St_f = 0.4$.

## 3.5 Lock-in condition

As explained in section 1, in the absence of deformation of the splitter plate, lock-in (phase-locking, synchronization) is a phenomenon in which the vortex shedding frequency changes (and locks) to match the applied flow perturbation frequency. Extending this to deformable plates, lock-in occurs if beating does not occur in the $Y_{tip}$ signal and the plate frequency synchronizes to the forcing frequency. This situation leads to maximum plate deformation. Fig. 11A summarizes the simulations performed, plotting beating and lock-in cases as a function of forcing frequency ($St_f$) for different applied forcing flow amplitudes ($K$). The natural frequencies of the plate for the first three modes ($St_{n1}$, $St_{n2}$ and $St_{n3}$) are also plotted. The lock-in occurs for the flow amplitude, $K \geq 0.4$ and at a forcing frequency, $St_f = 0.4$, twice of the splitter plate oscillation frequency as well as its natural frequency in the second mode, i.e., $St_f \cong 2St_p \cong 2St_{n2}$. The natural frequencies are calculated using eq. 14 and are listed for first three modes in Table 6. Note that the simulations at $St_f = 0.35$ and 0.45 at different flow amplitudes show beating patterns. In Fig. 11B, the increase in the plate displacement ($Y_{tip}$) at lock-in is quantified by plotting percentage change in it with respect to the steady inflow (eq. 18) as a function of the forcing flow amplitudes. $\Delta \eta$ for $Y_{tip}$ linearly increases with the flow amplitude and reaches to around 35%, at $K = 1$, as plotted in Fig. 11B. Superimposed shapes of the deformed plate at different time instances are shown in the inset of Fig. 11B for a typical lock-in case.

The effects of the lock-in on the plate motion are quantified by comparing the phase plane plot of the trajectory of the plate tip ($X_{tip}$, $Y_{tip}$) for the steady inflow and pulsatile inflow ($St_f =0.4$, $K=0.4$) in Fig. 12. Around 20 plate oscillation cycles are plotted for both cases after plate reaches self-sustained oscillation with plateau displacement. Results shows that the plate oscillates about a mean position in both cases and $Y_{tip}$ is around 15% larger in pulsatile flow case due to lock-in condition. The influence of lock-in on the associated wake structures is shown by qualitatively comparing vorticity contours at different instances in a typical cycle in insets in Fig. 12. The right insets for pulsatile inflow case show that the shear layers at the top and bottom of the plate roll up to form strong positive and negative vortices at the instance of maximum plate deformation.

In order to quantify the effect of structure-fluid density ratio ($\rho_s$) on the lock-in and beating conditions, we performed additional simulations by varying it to $\rho_s = 5$ and keeping all other parameters same ($E = 1400$, $Re = 100$, $L = 3.5D$) in the cases considered in section 3.3 and 3.4. The beating and lock-in cases are plotted in Fig 13A for several values of forcing frequency ($St_f$) and forcing amplitude ($K$). The lock-in occurs for all flow amplitudes, $K > 0$ and at a forcing frequency, $St_f = 0.46$, roughly twice of the natural frequency in the second mode (eq. 14), i.e., $St_f \cong 2St_{n2}$. Superimposed shapes of the deformed plate at different time instances are shown in the inset of Fig. 13A for a typical lock-in case.

Finally, we vary the plate length to $L = D$ for plotting the lock-in and beating cases. The other parameters are kept same as in sections 3.3 and 3.4 ($E = 1400$, $\rho_s = 10$). The Reynolds number, $Re = 200$ is used in these simulations since the plate displacement for $L = D$, $Re = 100$ is too small (on the order of 0.01) to evaluate the lock-in and beating conditions. The lock-in and beating cases are plotted in Fig 13B as function of forcing frequency ($St_f$) and forcing amplitude ($K$). The inset in the figure shows the superimposed shapes of the deformed plate at different time instances. The lock-in occurs for flow amplitude, $K \geq 0.2$ and at a forcing frequency, $St_f = 0.76$, twice of the natural frequency in the first mode (eq. 14) i.e., $St_f \cong 2St_{n1}$. Therefore, in all three cases considered (Fig. 11A, Fig 13A and Fig. 13B), the lock-in occurs when the applied oscillation frequency is around twice of the natural frequency in a particular mode. The mode of the natural frequency depends on the plate length, as discussed in section 3.2.

## 4 Conclusions

The effect of the pulsatile inflow on the flow-induced deformation of an elastic plate inside a channel is simulated numerically by combining a sharp-interface immersed boundary method flow solver and an open-source finite-element based structural solver. The coupling is accomplished using an implicit iterative approach to improve the stability properties of the combined solver. The plate exhibits small deformations initially, asymptoting to a periodic self-sustained oscillation at longer times for the steady inflow. In the case of the pulsatile inflow, the plate experiences strong forcing from vortices that form from the separating shear layers from the cylinder and subsequently advect downstream over the surfaces of the plate. Despite the tendency of the vortices to form and shed symmetrically because of the applied longitudinal forcing, the coupling with the allowable cross-stream oscillation mode of the plate leads to substantially increased cross-stream oscillation amplitude relative to the unforced case in general. The maximum plate displacement is observed when the applied oscillation frequency is twice the natural plate oscillation frequency in a particular mode ($St_f \cong 2St_{ni}$), corresponding to the resonant or lock-in case. The mode of the natural frequency depends upon the plate length. For applied frequencies away from this condition, beating is observed due to the superposition of the applied and natural oscillatory signals. The plate deformation response, and drag of the plate and its components, are quantified for forcing flow amplitudes $K \leq 1$, and for forcing frequencies $St_f \leq 1$. The total drag on the plate is found to be significantly larger relative to the steady inflow, at forcing frequencies equal or larger than lock-in frequency, at a given flow amplitude. For lock-in cases, the plate displacement, total drag, pressure as well as skin friction drag increases with the forcing flow amplitude.

# 5 Acknowledgements

R.B. gratefully acknowledges financial support from the Department of Science and Technology, New Delhi through its fast track scheme for young scientists. This work was also partially supported by an internal grant from Industrial Research and Consultancy Centre (IRCC), IIT Bombay. A.K. was supported by Prime Minister Fellowship for doctoral research sponsored by Confederation of Indian Industry, New Delhi and Forbes Marshall Inc, Pune, India. We thank Prof. John Sheridan at Monash University for useful discussions as well as two anonymous reviewers for useful comments.

# 7 Tables

Table 1: Comparison between flow quantities for steady flow past a circular cylinder at $Re=100$ with published data [35, 36].

| Flow quantities | Present work | Li et.al. [35] | Behr et.al. [36] |
|---|---|---|---|
| Maximum lift coefficient | 0.3354 | 0.3630 | 0.3743 |
| Average drag coefficient | 1.3801 | 1.3415 | 1.3836 |
| Strouhal number | 0.1639 | 0.1678 | 0.1641 |

Table 2: Comparison of dominant frequencies in lift signals at different forcing amplitude ($A_f$) for pulsatile flow past a circular cylinder at $Re = 100$ with those reported by Li et al. [35].

| Cases | Present work | | Li et.al. [35] | |
|---|---|---|---|---|
| | First frequency | Second frequency | First frequency | Second frequency |
| $A_f = 0.05$ | 0.1645 | 0.3215 | 0.1678 | 0.3357 |
| $A_f = 0.08$ | 0.1647 | 0.3312 | 0.1602 | 0.3281 |
| $A_f = 0.1$ | 0.1639 | 0.3279 | 0.1526 | 0.3204 |
| $A_f = 0.4$ | 0.1647 | 0.3246 | 0.1678 | 0.3357 |

Table 3: Comparison between computed and published results reported for the FSI benchmark.

| Study | Plate oscillation frequency, $St_p$ | Plate maximum displacement, $Y_{tip}$ |
|---|---|---|
| Present work | 0.19 | 0.83 |
| Turek and Hron [14] | 0.19 | 0.83 |
| Bhardwaj and Mittal [15] | 0.19 | 0.92 |
| Lee and You [16] | 0.19 | 0.85 |
| Furquan and Mittal [18] | 0.19 | 0.79 |
| Tian et al. [37] | 0.19 | 0.78 |

Table 4: Grid size convergence study. Error in the maximum plate tip deflection ($Y_{tip}$) for different grids with respect to the finest grid examined.

| Cases | Grid | $\Delta x_{min}$ | $\Delta y_{min}$ | Relative error in $Y_{tip}$ with respect to case 3 |
|---|---|---|---|---|
| 1 | 256 × 160 | 0.03 | 0.020 | -0.075 % |
| 2 | 384 × 160 | 0.02 | 0.020 | -0.002% |
| 3 | 480 × 224 | 0.015 | 0.014 | - |

Table 5: Domain size independence study. Error in the maximum plate tip deflection ($Y_{tip}$) for different domain sizes with respect to the longest domain size examined.

| Cases | Domain size | Relative error in $Y_{tip}$ with respect to case 4 |
|---|---|---|
| 1 | $20D \times 4.1D$ | 0.042% |
| 2 | $25D \times 4.1D$ | 0.021% |
| 3 | $30D \times 4.1D$ | 0.010% |
| 4 | $40D \times 4.1D$ | - |

Table 6: Simulation results for different cases considered for steady inflow. The Reynolds number ($Re$), structure-fluid density ratio ($\rho_s$) and plate length ($L$) are varied. The Young's Modulus and plate thickness are 1400 and 0.2, respectively, in all cases. The natural frequencies of the plate in first three modes ($St_{ni}$) calculated using eq. 14 are also listed for all cases.

| Cases | $Re$ | $\rho_s$ | $L$ | $Y_{tip}$ | $St_p$ | $St_{n1}$ | $St_{n2}$ | $St_{n3}$ | Remarks |
|---|---|---|---|---|---|---|---|---|---|
| 1 | 100 | 10 | 3.5 | 0.83 | 0.19 | 0.03 | 0.20 | 0.55 | $St_p \cong St_{n2}$ |
| 2 | 200 | 10 | 3.5 | 0.89 | 0.20 | 0.03 | 0.20 | 0.55 | $St_p \cong St_{n2}$ |
| 3 | 100 | 5 | 3.5 | 0.45 | 0.23 | 0.04 | 0.28 | 0.78 | $St_p \cong St_{n2}$ |
| 4 | 100 | 10 | 1 | 0.37 | 0.35 | 0.38 | 2.40 | 6.71 | $St_p \cong St_{n1}$ |

# 8 Figures

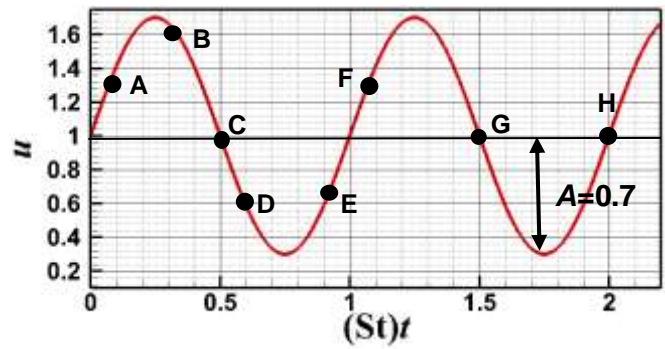

(a)

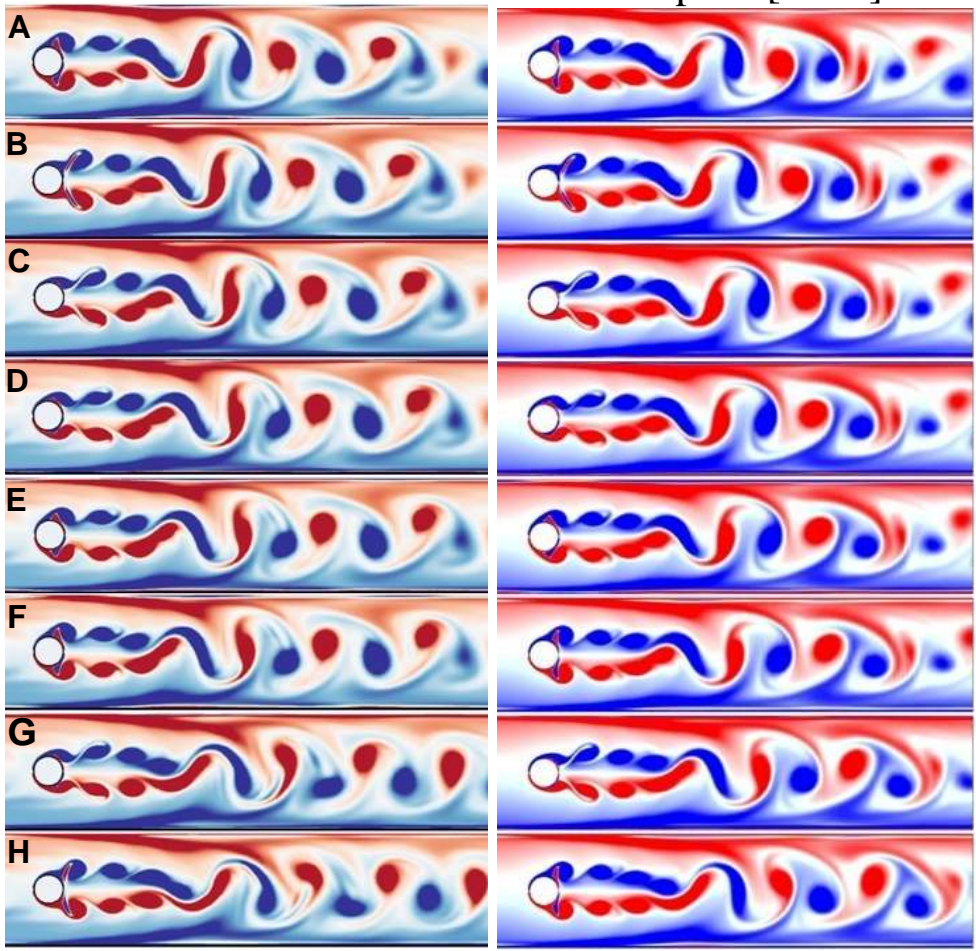

(b)

**Figure 1:** Validation against the wake structure of Al-Sumaily and Thompson [34] for the pulsatile flow past a circular cylinder placed in a horizontal channel. Left: present wake

vorticity predictions, Right: vorticity fields from Al-Sumaily and Thompson (Adapted with permission from [34]). In both cases *St*= 0.8, *Re* = 250 and $A_f$ =0.7.

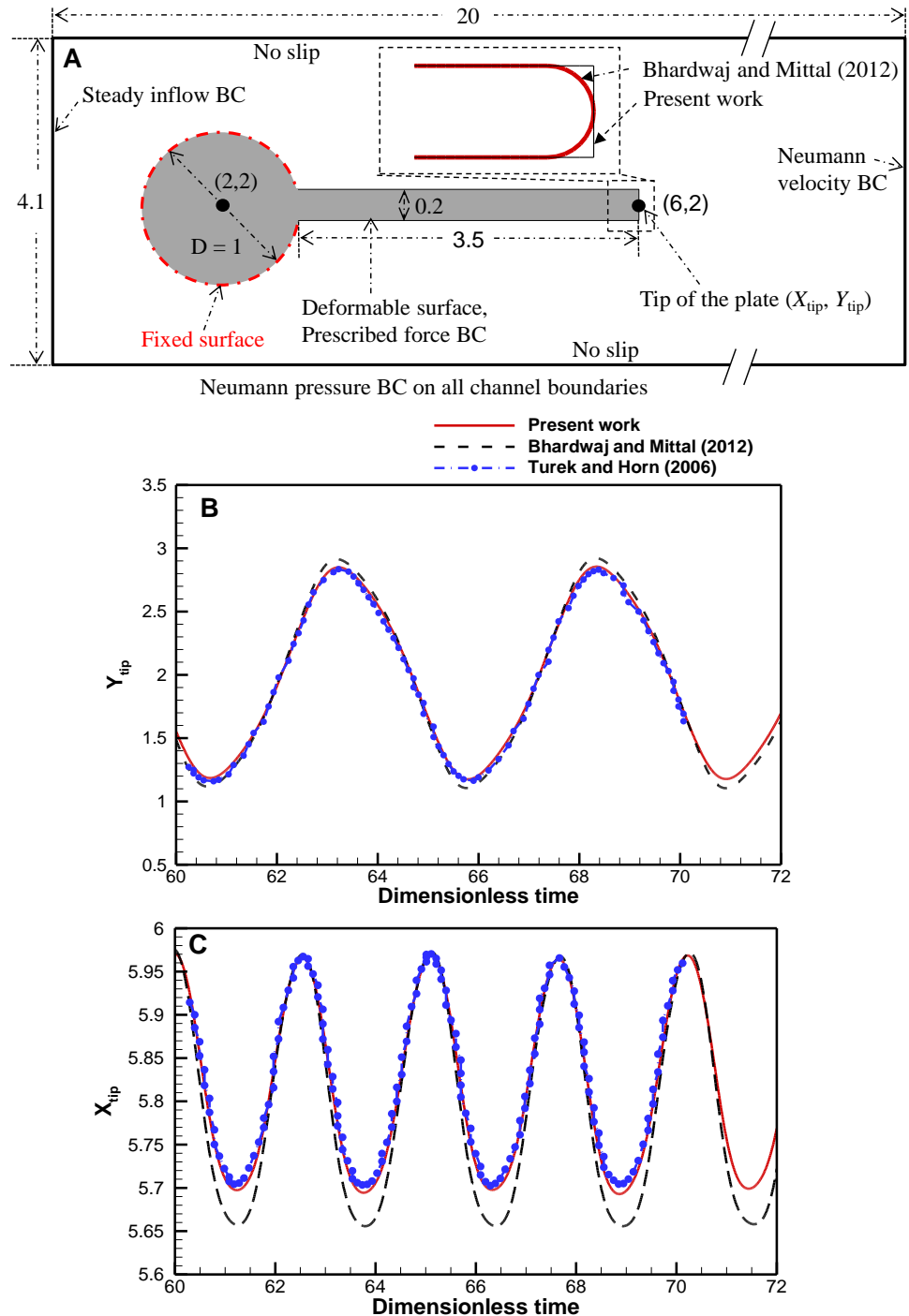

Figure 2: (A) Schematic of the computational domain with boundary conditions (BC) considered for the FSI benchmark. The benchmark was first proposed by Turek and Hron [14] and later considered by Bhardwaj and Mittal [15]. (B) Comparison of computed time-variation

of $Y_{tip}$ position with published results (C) Comparison of computed time-variation of $X_{tip}$ with the published results.

**(A)** $L = 3.5D$

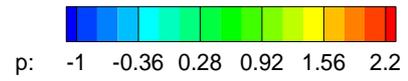

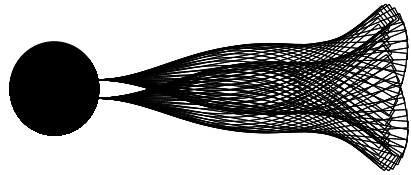
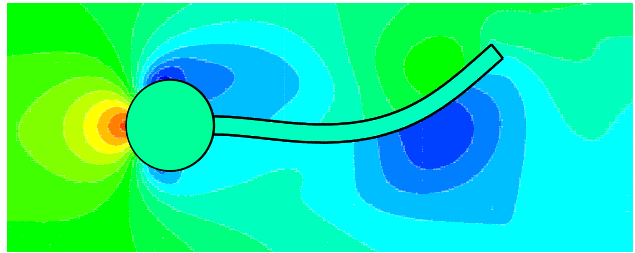

**(B)** $L = D$

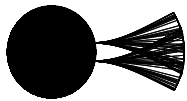
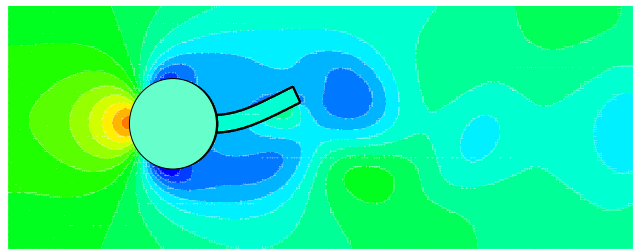

Figure 3: Superimposed deformed shapes of the plate at several time instances for the plate length (left) and pressure contours (right) for steady inflow at the inlet for different plate lengths. (A) $L = 3.5D$ (B) $L = D$. Other parameters used in both cases are $E = 1400$, $Re = 100$ and $\rho_s = 10$.

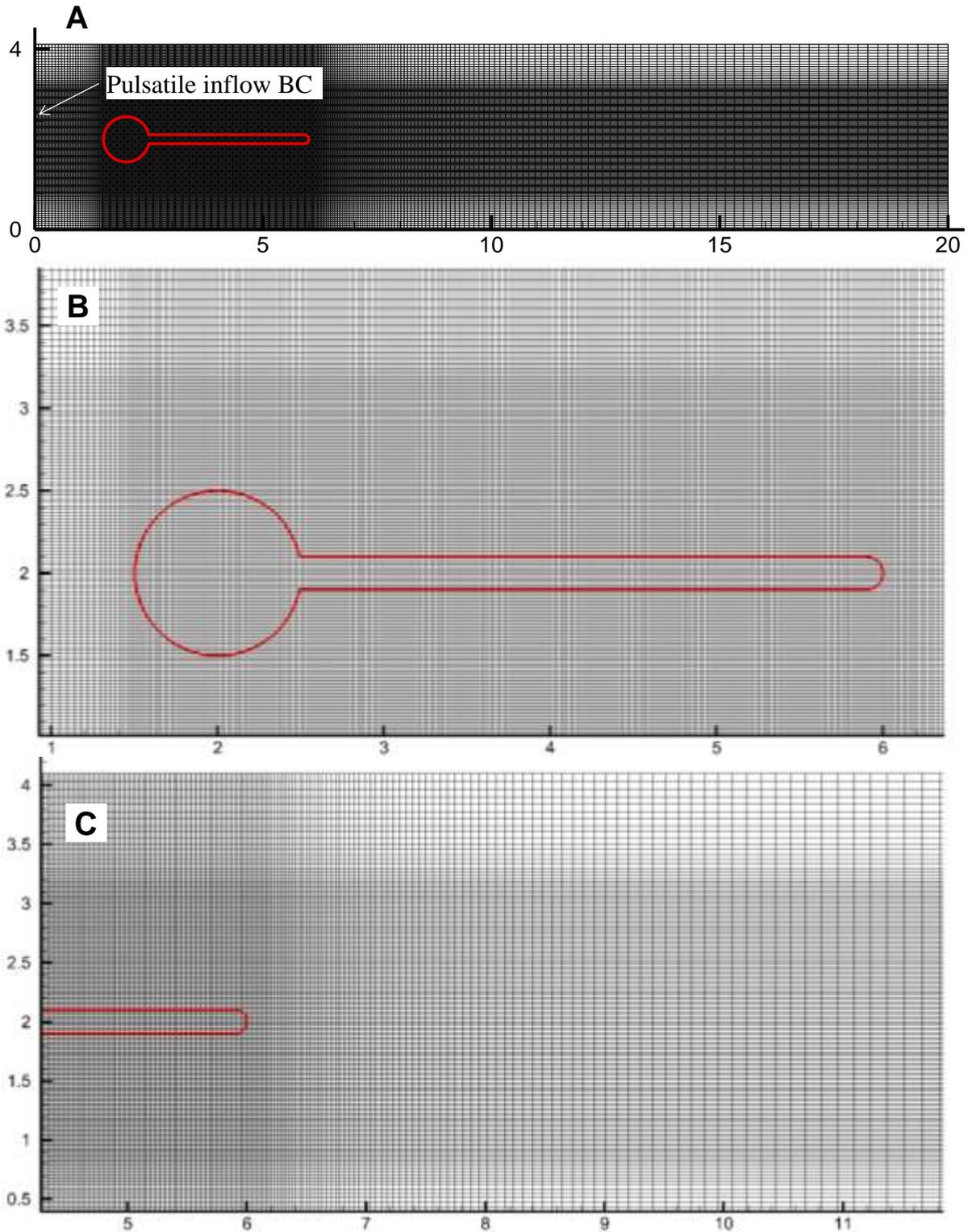

Figure 4: (A) Non-uniform Cartesian grid in the computational domain. (B) Zoomed in view of the grid in the vicinity of the immersed boundary. Uniform grid is used in the region in which the plate is expected to move and non-uniform grid stretching is used from this region to the wall. The immersed boundary (fluid-structure interface) is shown in red. (C) Zoomed in view of the grid in the downstream with grid stretching used away from the tip of the plate.

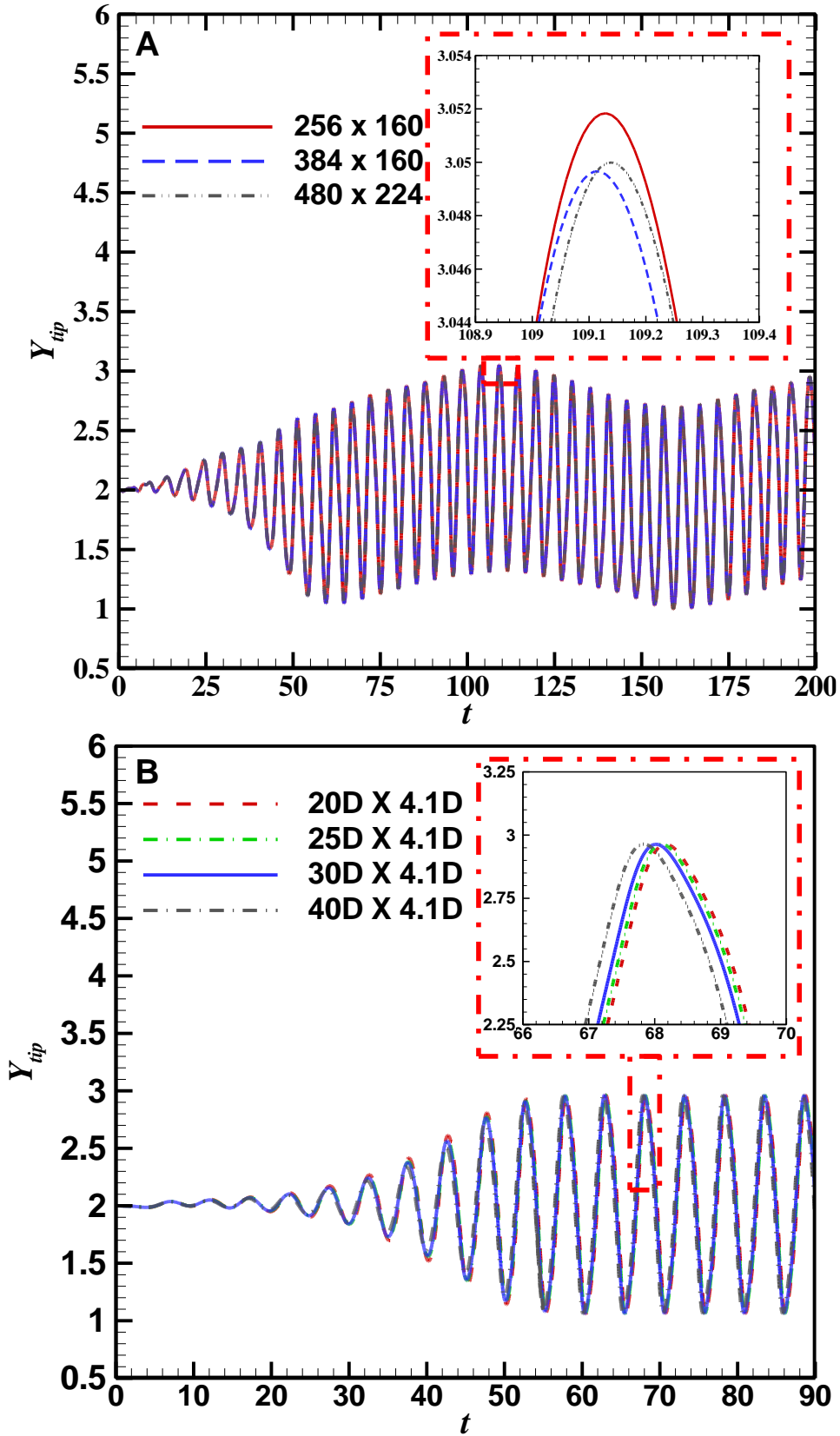

Figure 5: (A) Grid size convergence study: comparison of the time-varying cross-stream displacement of the plate tip ($Y_{\text{tip}}$) as a function of grid resolution of the immersed boundary method solver. (B) Domain size independence study: comparison of the time-varying cross-stream displacement of the plate tip ($Y_{\text{tip}}$) for different domains.

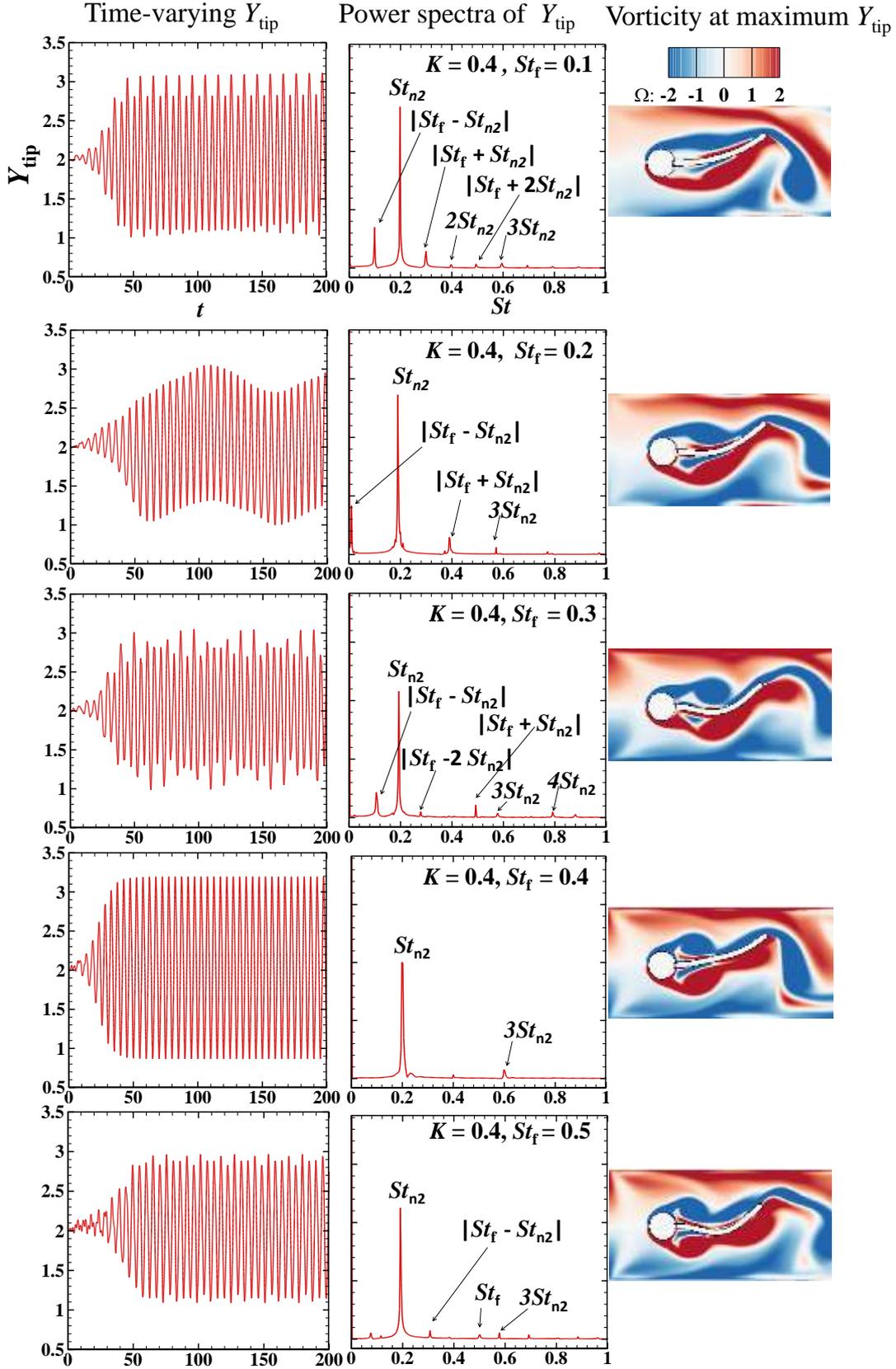

**Figure 6A:** Time-varying tip displacement varying with amplitude ($K$) and frequency ($St_f$). Note that $K = 0$, $St_f = 0$ corresponds to non-pulsatile flow. Power spectra of $Y_{tip}$ displacement of elastic plate (middle figure). Vorticity distribution of pulsatile flow (right).

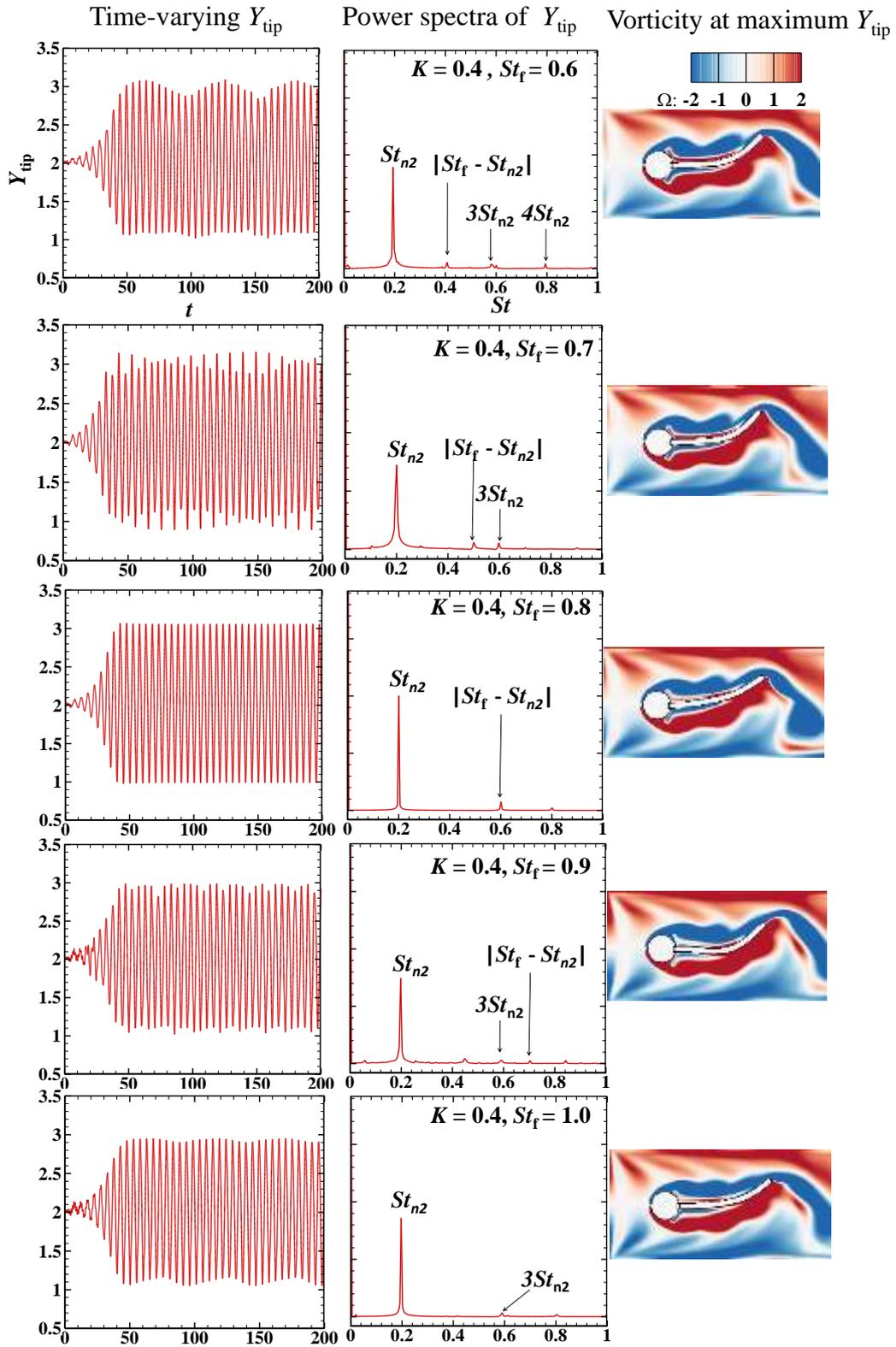

Figure 6B: Continuation of (A)

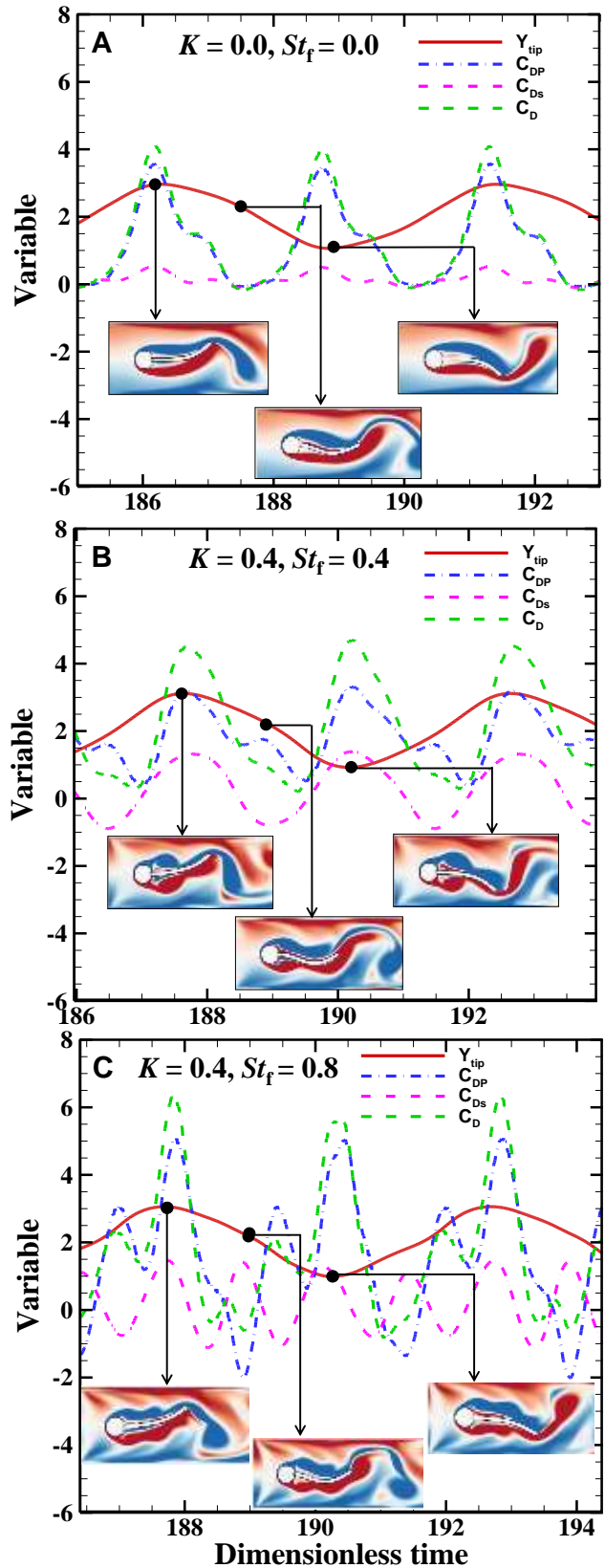

Figure 7: Comparison among time-varying displacement of the tip of the plate ($Y_{\text{tip}}$), pressure drag coefficient ($C_{\text{DP}}$), skin friction drag coefficient ($C_{\text{DS}}$) and total drag coefficient ($C_{\text{D}}$). (A) Steady inflow, $K = 0.0$, $St_{\text{f}} = 0.0$ (B) Pulsatile inflow, $K = 0.4$, $St_{\text{f}} = 0.4$, (C) Pulsatile inflow, $K = 0.4$, $St_{\text{f}} = 0.8$. Vorticity distribution at different time-instances is shown in insets and the vorticity scale is same as in Fig 6 (third column).

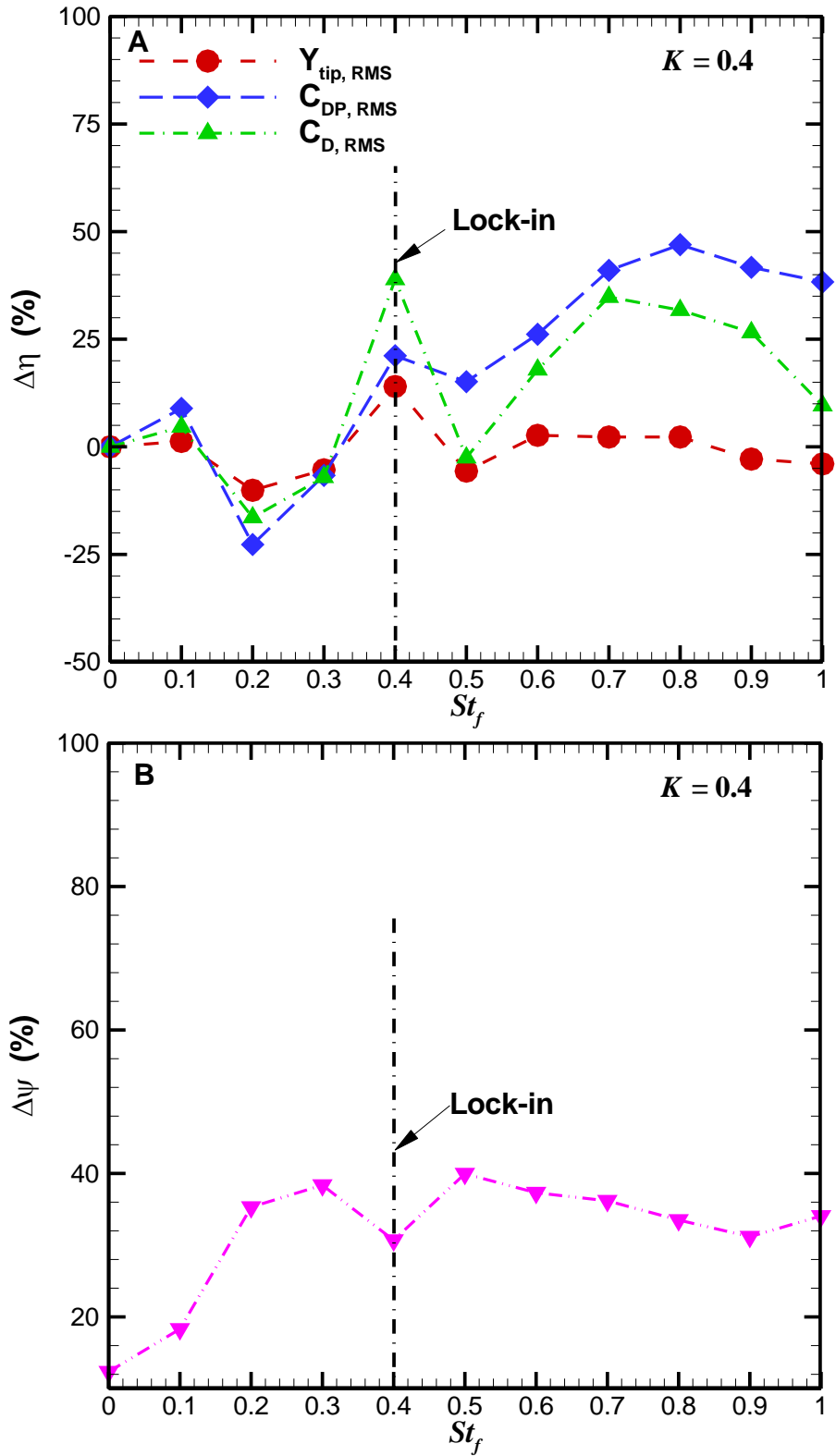

Figure 8: (A) Percentage change in RMS values of deformation of plate ($Y_{tip}$), pressure drag coefficient ($C_{DP}$) and total drag coefficient ($C_D$), relative to steady inflow for several pulsatile inflow frequencies ($St_f$) at constant inflow amplitude ($K = 0.4$). (B) Contribution of the RMS skin friction drag with respect to the RMS total drag.

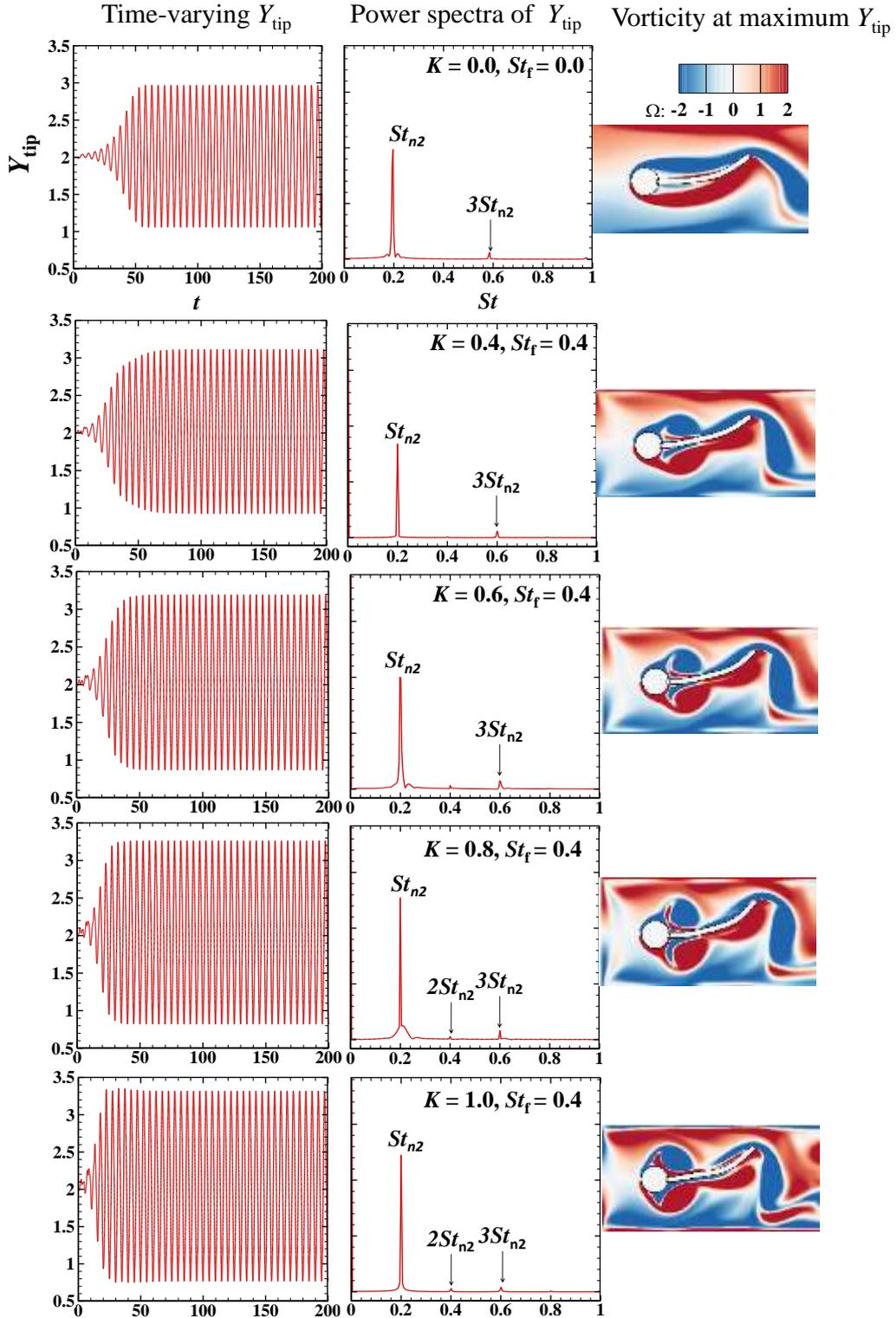

**Figure 9:** Time-varying $Y$-displacement of the tip of the plate varying with amplitude ($K$) for constant frequency $St_f = 0.4$. Note that $K = 0$, $St_f = 0$ corresponds to steady flow at the inlet. Effect of amplitude on elastic plate (left figure). Power spectra of $Y_{tip}$ displacement of elastic plate (middle figure). Vorticity distribution of pulsatile flow (right).

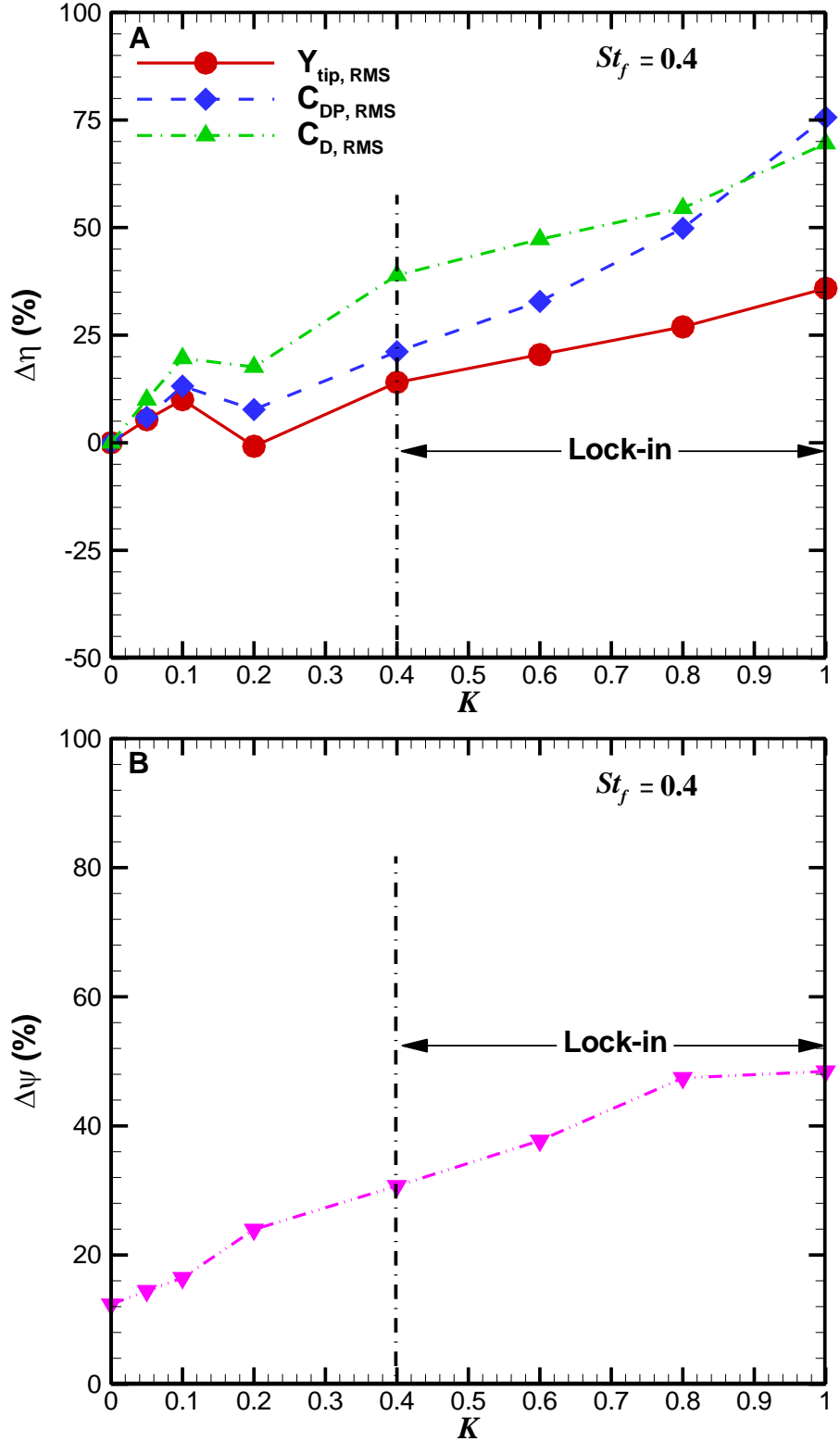

Figure 10: Percentage change in RMS values of deformation of plate ($Y_{tip}$), pressure drag coefficient ($C_{DP}$) and total drag coefficient ($C_D$), relative to results for steady inflow, as a function of pulsatile inflow amplitude ($K$). The inflow frequency was fixed at $St_f = 0.4$. (B) Contribution of the RMS skin friction drag with respect to RMS total drag.

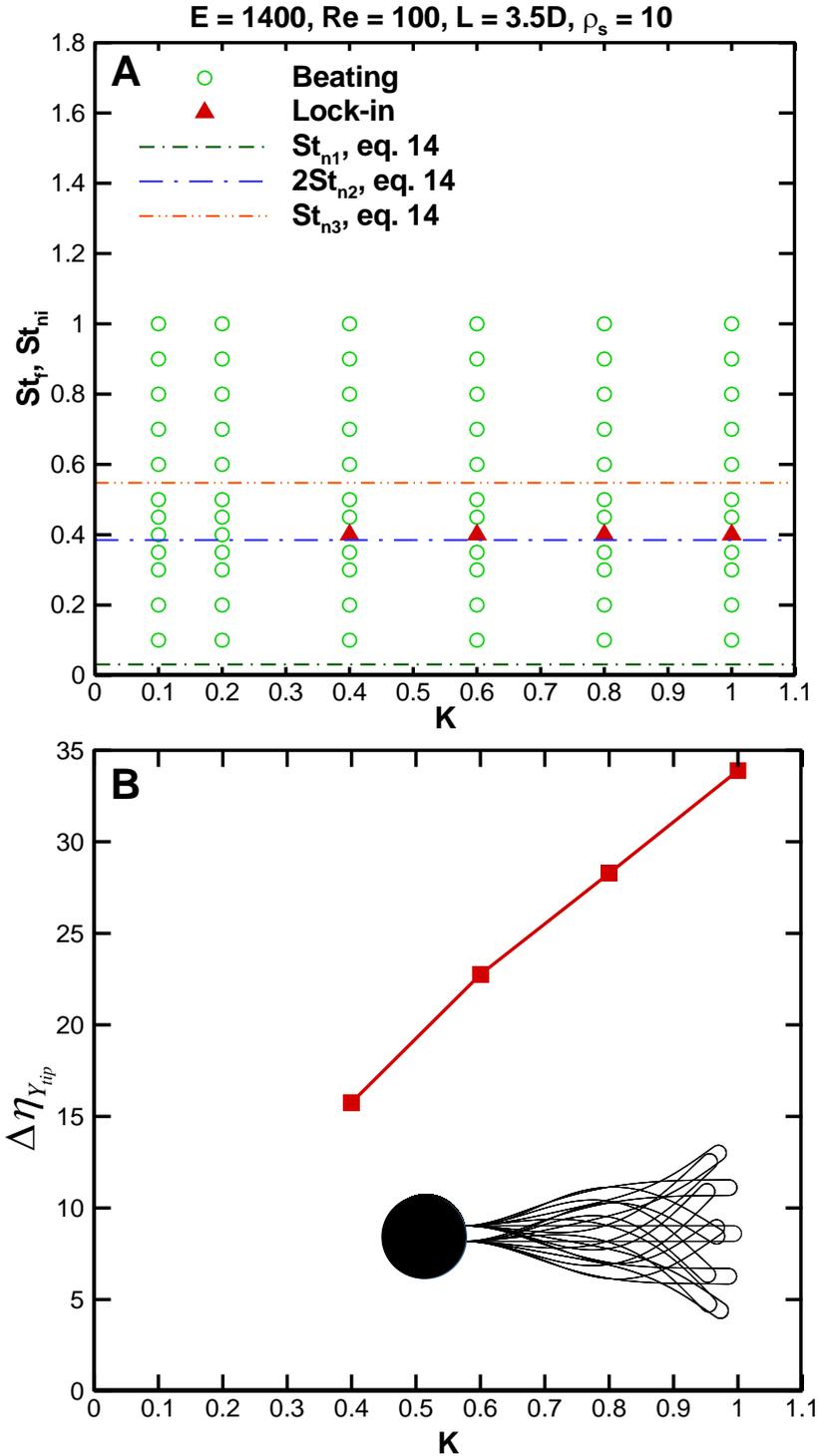

**Figure 11:** (A) Lock-in and beating plotted as a function of forcing amplitude ($K$) and forcing frequency ($St_f$). The first three modes of natural frequency of the plate ($St_{n1}$, $St_{n2}$ and $St_{n3}$) are plotted as lines. The lock-in occurs if the forcing frequency is twice of the second mode of natural frequency ($St_f \cong 2St_{n2}$). (B) Percentage increase in the tip displacement of the plate at lock-in condition with respect to steady inflow. The maximum plate oscillation amplitude occurs at lock-in, when the forcing frequency is twice the plate oscillation frequency. The different points (filled squares) correspond to different forcing amplitudes. The inset shows superimposed deformed shapes of the plate at several time instances for a typical lock-in case.

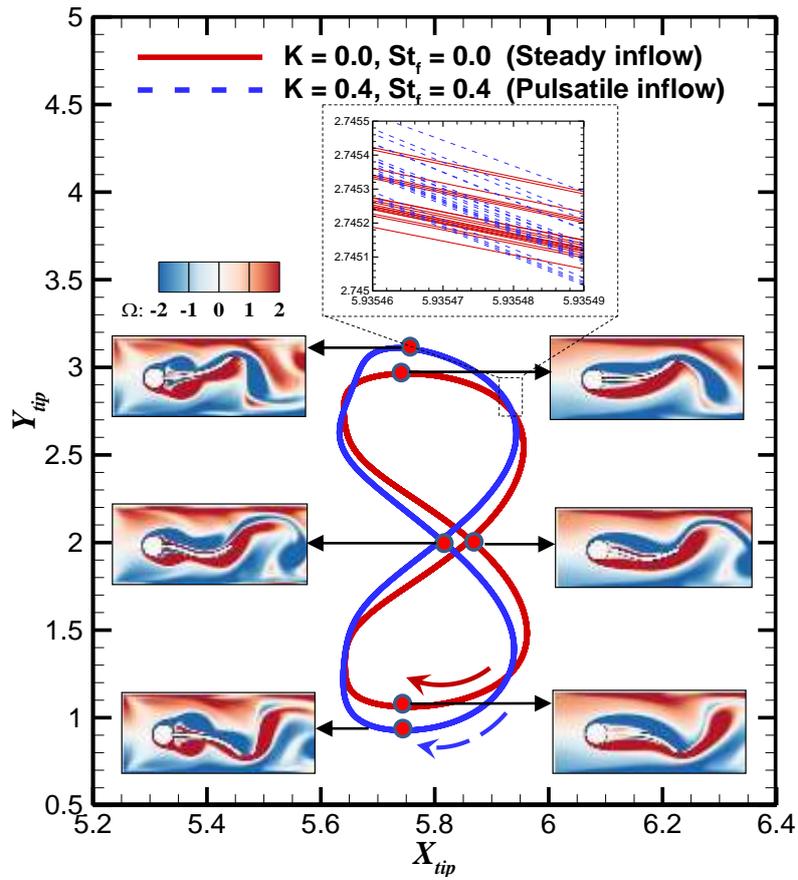

Figure 12: Comparison between the phase plane plots of the trajectories of the plate tip are for pulsatile and steady inflow. Around 20 plate oscillation cycles are plotted for both cases, shown in the top inset. Insets on the left and right show the corresponding vorticity contours at critical points in the cycle.

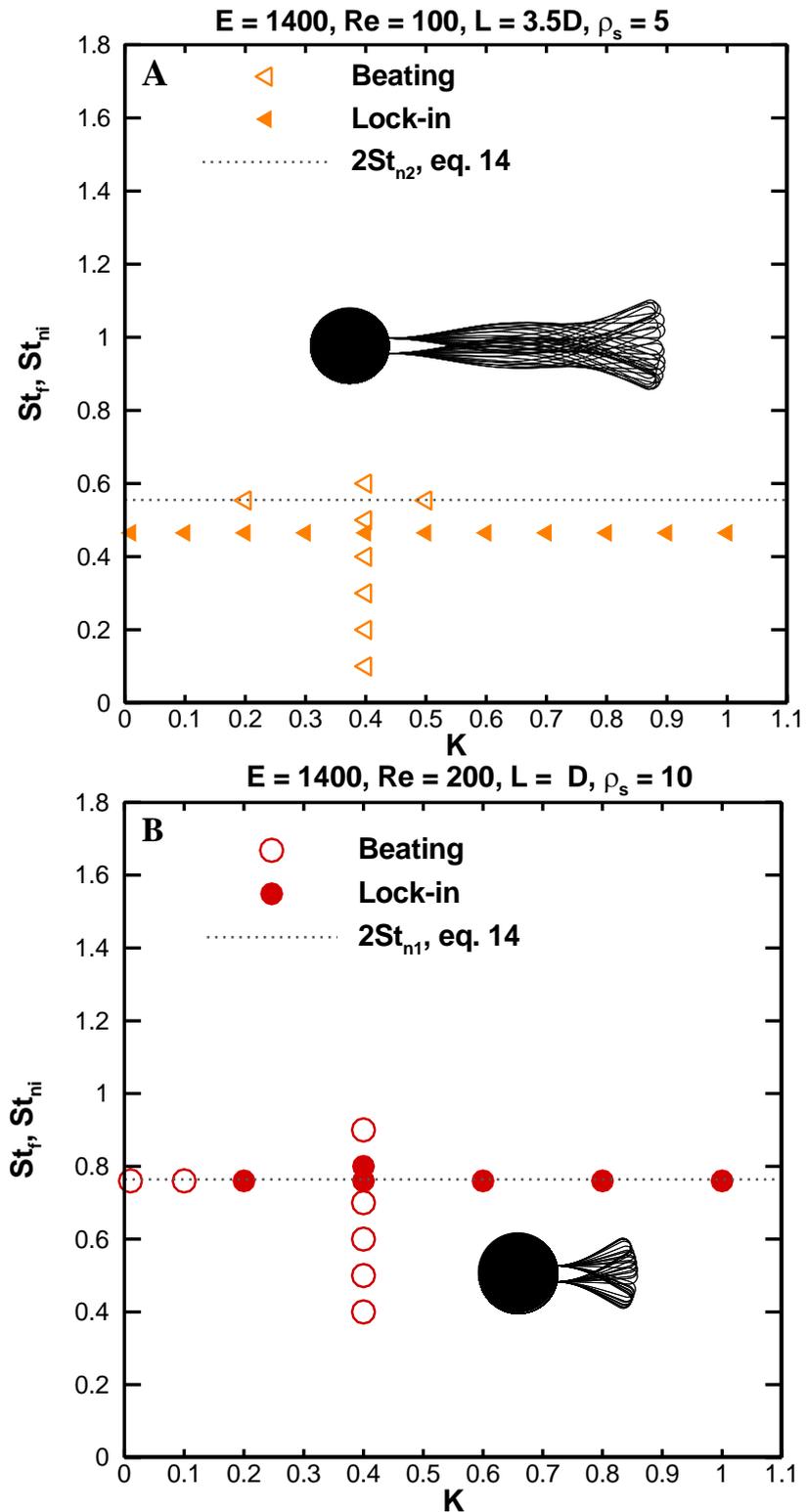

Figure 13: Lock-in and beating plotted as a function of forcing amplitude ($K$) and forcing frequency ($St_f$). (A) The structure-fluid density ratio is varied to $\rho_s = 5$. (B) The plate length is varied to, $L = D$. The insets show superimposed deformed shapes of the plate at several time instances for a typical lock-in cases. The second mode of natural frequency ($St_{n2}$) is plotted as broken lines. The lock-in occurs if the forcing frequency is twice of natural frequency in second mode ($St_f \cong 2St_{n2}$).

1
2
3
4
5
6
7
8
9
10
11
12
13
14
13
16
17
18
19
20
21
22
23
24
25
26
13
28
22
23
24
32
33
34
35
36
31

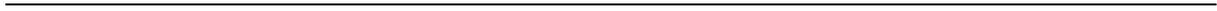
37